\documentclass[aps,twocolumn,superscriptaddress,showpacs,pre]{revtex4}
\usepackage{graphicx}

\begin{document}
\title{Transverse power flow reversing of confined waves in extreme nonlinear metamaterials}

\author{A. Ciattoni}
\affiliation{Consiglio Nazionale delle Ricerche, CASTI Regional Lab 67100 L'Aquila, Italy and Dipartimento di Fisica, Universit\`{a} dell'Aquila,
67100 L'Aquila, Italy}

\author{C. Rizza}
\affiliation{Dipartimento di Ingegneria Elettrica e dell'Informazione, Universit\`{a} dell'Aquila, 67100 Monteluco di Roio (L'Aquila), Italy}

\author{E. Palange}
\affiliation{Dipartimento di Ingegneria Elettrica e dell'Informazione, Universit\`{a} dell'Aquila, 67100 Monteluco di Roio (L'Aquila), Italy}

\date{\today}

\begin{abstract}
We theoretically prove that electromagnetic beams propagating through a nonlinear cubic metamaterial can exhibit a power flow whose direction
reverses its sign along the transverse profile. This effect is peculiar of the hitherto unexplored extreme nonlinear regime where the nonlinear
response is comparable or even greater than the linear contribution, a condition achievable even at relatively small intensities. We propose a
possible metamaterial structure able to support the extreme conditions where the polarization cubic nonlinear contribution does not act as a mere
perturbation of the linear part.
\end{abstract}

\maketitle
The ability of manufacturing metamaterials with prescribed and anomalous values of permittivity $\epsilon$ and permeability $\mu$ has triggered an
intense research effort aimed at investigating novel regimes of linear electromagnetic propagation and suitable configurations have been devised for
observing remarkable effects such as, for example, superlensing \cite{Pendr1,FangLe}, optical cloaking \cite{Pendr2,Schuri}, guiding of nanometric
optical beams \cite{Takaha} and photonic circuits \cite{Silver,Enghet}. In the nonlinear realm, the nonlinear properties of left-handed metamaterials
have been investigated \cite{Zharov} together with various soliton manifestations \cite{Shadri,Lazari,Zharoa}. Propagation in metamaterials
exhibiting cubic nonlinear response has also been considered \cite{Liubar,Huwenz} and, for ultra-short pulse nonlinear dynamics, it has been
suggested that metamaterial linear property tailoring allows the observation of different nonlinear regimes \cite{Scalor}.

In this Letter we show that a metamaterial with a very small linear dielectric constant and exhibiting a nonlinear cubic response is able to support
nonlinear guided waves whose Poynting vector has the very peculiar property of being parallel and anti-parallel to the propagation direction in
different transverse portion of the field. This novel phenomenology is a consequence of the fact that, since the metamaterial linear dielectric
permittivity can be arbitrary small, the nonlinear contribution to the dielectric response can easily (i.e. at low intensities) be made comparable or
greater than the linear part so that, the sign of the overall dielectric response can be different for different intensities. In the presence of an
electromagnetic beam this implies that conditions can be found so that the effective dielectric response has different signs on the propagation axis
and at its lateral sides. Therefore the transverse reversing of the power flow is understood since, for a monochromatic Transverse Magnetic (TM)
field mainly propagating along a given direction, the Poynting vector globally lies along the same mean propagation direction and its sign coincides
with that of the total effective dielectric constant. In order to discuss this effect on a feasible situation, we consider TM electromagnetic
propagation in a defocusing nonlinear cubic metamaterial and we analytically obtain a class of nonlinear guided waves exhibiting the aforementioned
transverse power flow reversing. It is remarkable that the power flow reversing effect can be observed even at very low intensities since it is a
consequence of the interplay between the linear and nonlinear contributions to the dielectric response, the former being very small in the considered
metamaterials and the latter being proportional to the intensity. To the best of our knowledge, the transverse power flow reversing predicted in the
present Letter is the first example of effects characterizing a novel extreme nonlinear regime where the nonlinear contribution to the medium
polarization does not play the role of a mere perturbation to the linear part. The question naturally arises as to whether a medium exists or can be
conceived where the range of electromagnetic intensities, for which its nonlinear response is purely cubic, is so large to produce a huge nonlinear
response. At first one may reject this possibility since the cubic nonlinear response generally arises from a perturbative description of
radiation-matter interaction so that the nonlinear polarization necessarily is a small correction to the linear part. However, exploiting the
availability of metamaterials with somehow prescribed values of the dielectric permittivity, we propose that in a suitable sub-wavelength layered
structure, consisting of alternating slabs of a metamaterial with negative dielectric constant and a standard nonlinear cubic medium, the effective
electromagnetic response is purely cubic in an intensity range where the nonlinear cubic term can exceed the linear contribution.
\begin{figure}
\includegraphics[width=0.5\textwidth]{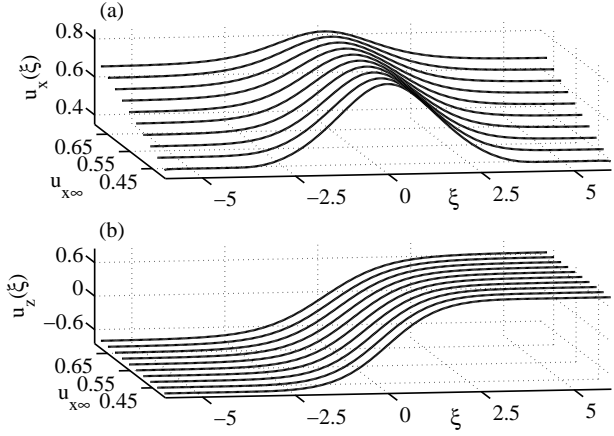}
\caption{Nonlinear guided waves transverse profile of $u_x$ (panel (a)) and of $u_z$ (panel (b)) at different values of $u_{x\infty}$ in the range of
Eq.(\ref{exist}), for $\gamma = 0.5$.}
\end{figure}

Consider a monochromatic electromagnetic field (whose time variation is assumed to be $e^{-i \omega t}$, where $\omega$ is the angular frequency)
propagating through a nonlinear metamaterial characterized by the constitutive relations (holding for the field complex amplitudes)
\begin{eqnarray} \label{consti}
{\bf D} &=& \epsilon_0 \epsilon {\bf E} - \epsilon_0 \chi [({\bf E} \cdot {\bf E}^*) {\bf E} + \gamma ({\bf E} \cdot {\bf E}) {\bf E}^* ], \nonumber \\
{\bf B} &=& \mu_0 \mu {\bf H},
\end{eqnarray}
where $\epsilon>0$ and $\mu>0$ are the linear permittivity and permeability, respectively, whereas $\chi>0$ and $0<\gamma<1$ are the parameters
characterizing the cubic defocusing nonlinear response. We focus our attention on transverse magnetic (TM) nonlinear guided waves propagating along
the $z-$ axis of the form
\begin{eqnarray} \label{TM}
{\bf E}(x,z) &=& e^{i \beta \zeta} \sqrt{\frac{\epsilon}{\chi}} [u_x (\xi) \hat{\bf e}_x + i u_z (\xi) \hat{\bf e}_z ], \nonumber \\
{\bf H}(x,z) &=& e^{i \beta \zeta} \sqrt{\frac{\epsilon_0 \epsilon^2}{\mu_0 \mu \chi}} \left[\beta u_x(\xi) - \frac{du_z (\xi)}{d\xi} \right]
\hat{\bf e}_y
\end{eqnarray}
where $\xi = \sqrt{\epsilon \mu} (\omega/c) x$, $\zeta = \sqrt{\epsilon \mu} (\omega/c) z$ ($c$ is the speed of light in vacuum) are dimensionless
spatial coordinates, $\beta$ is a real dimensionless propagation constant and $u_x$ and $u_z$ are dimensionless electric field components.
Substituting the fields of Eqs.(\ref{TM}) into Maxwell equations $\nabla \times {\bf E} = i \omega {\bf B}$ and $\nabla \times {\bf H} = -i \omega
{\bf D}$ and using the constitutive relations of Eqs.(\ref{consti}) we get
\begin{eqnarray} \label{syst}
\beta \frac{d u_z}{d \xi} &=& \left[ (\beta^2-1)+(1+\gamma) u_x^2+(1-\gamma) u_z^2 \right] u_x, \nonumber \\
\frac{d^2 u_z}{d \xi^2} - \beta \frac{d u_x}{d \xi} &=& \left[ -1 + (1-\gamma) u_x^2 + (1+\gamma) u_z^2 \right] u_z
\end{eqnarray}
which is a system of ordinary differential equations fully characterizing the transverse profile of the considered nonlinear guided waves. Without
loss of generality we consider solutions of Eqs.(\ref{syst}) with definite parity where $u_x$ and $u_z$ are spatially even ($u_x(\xi) = u_x(-\xi)$)
and odd ($u_z(\xi) = - u_z(-\xi)$), respectively and, as a consequence, we adopt the boundary conditions $u_x(0)=u_{x0}$, $u_z(0)=0$ and
$u_x(+\infty)=u_{x\infty}$, $u_z(+\infty)=u_{z\infty}$. Since $u_x(\xi)$ and $u_z(\xi)$ have to asymptotically approach two constant values, their
first and second derivative vanish for $\xi \rightarrow +\infty$ so that, exploiting the boundary conditions, we require the right hand sides of
Eqs.(\ref{syst}) to vanish at $u_x=u_{x\infty}$ and $u_z=u_{z\infty}$. Therefore we obtain $\beta = \sqrt{2 \gamma (1-2u_{x \infty}^2)/(1+\gamma)}$
and $u_{z \infty} = \sqrt{[1-(1-\gamma) u_{x \infty}^2]/(1+\gamma)}$ from which we note that $u_{x \infty}^2<1/2$ is a necessary condition for the
existence of the considered nonlinear waves. In order to prove their existence, we exploit the fact that the system of Eqs.(\ref{syst}) is integrable
\cite{Ciatto} since it admits the first integral
\begin{eqnarray} \label{F}
F(u_x,u_z) &=& (\beta^2-1) u_x^2 - u_z^2 + \frac{1}{2}(u_x^4+u_z^4)+(1-\gamma) u_x^2 u_z^2 + \nonumber \\
           &-& \frac{1}{\beta^2}\left[(\beta^2-1)+(1+\gamma)u_x^2+(1-\gamma)u_z^2\right]^2 u_x^2
\end{eqnarray}
\begin{figure}
\includegraphics[width=0.5\textwidth]{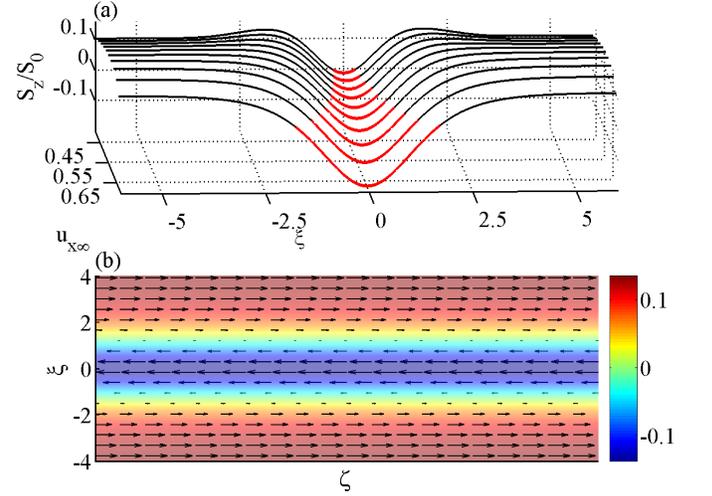}
\caption{(a) Profiles of the $z$- component of the Poynting vector (see Eq.(\ref{Poynt})) normalized with $S_0 = \sqrt{(\epsilon_0 \epsilon^3)/(4
\mu_0 \mu \chi^2)}$ corresponding to the fields reported in Fig.1. Each profile is characterized by an off-center positive part (black portion) and a
central negative part (red portion). (b) Plot of the field ${\bf S}/S_0$ (arrows) in the plane $(\xi,\zeta)$ corresponding to the nonlinear guided
wave with $u_{x\infty} = 0.65$ of Fig.(1). The color is related to the local value of $S_z / S_0$. Note the reversing of $\bf S$ along the transverse
$\xi$ axis.}
\end{figure}
or, in other words, the relation $\frac{d}{d \xi} F(u_x(\xi),u_z(\xi)) =0$ holds for any solution $u_x(\xi),u_z(\xi)$ of Eqs.(\ref{syst}). Evidently,
after substituting the obtained $\beta$ and $u_{z\infty}$ into Eq.(\ref{F}), $F$ has a stationary point at $(u_{x\infty},u_{z\infty})$ and the guided
waves are represented by curves of constat $F$ in the plane $(u_x,u_z)$ joining $(u_{x0},0)$ to the stationary point. Therefore, requiring that $F$
has a saddle point at $(u_{x\infty},u_{z\infty})$ and exploiting the above necessary condition, we conclude that the considered nonlinear waves exist
in the range
\begin{equation} \label{exist}
\sqrt{\frac{1}{2+  \frac{(\gamma+1)^2}{\gamma}}} < u_{x\infty} < \sqrt{\frac{1}{2}}
\end{equation}
which is always not empty since $\gamma >0$. In addition the relation $F(u_{x0},0)=F(u_{x\infty},u_{z\infty})$ yields the possible values of
$u_x(0)=u_{x0}$ corresponding to the asymptotical value $u_{x}(+\infty)=u_{x\infty}$. In Fig.1 we plot the profiles of $u_x$ and $u_z$ corresponding
to different values of $u_{x\infty}$, spanning the range of Eq.(\ref{exist}), for $\gamma = 0.5$ obtained by numerically integrating Eqs.(\ref{syst})
with the above boundary conditions. The power flow carried by these waves is described by the time-average Poynting vector ${\bf S} =
\frac{1}{2}Re({\bf E} \times {\bf H}^*)$ which, exploiting Eqs.(\ref{TM}) and the first of Eq.(\ref{syst}), becomes
\begin{equation} \label{Poynt}
{\bf S} = \sqrt{\frac{\epsilon_0 \epsilon^3}{4 \mu_0 \mu \chi^2}} \frac{1}{\beta}
          \left\{ 1 - \left[(u_x^2 + u_z^2)+\gamma(u_x^2 - u_z^2)\right] \right\} u_x^2 \hat{\bf e}_z
\end{equation}
i.e., for the considered waves, is purely along the $z-$ axis. In Fig.2(a) we plot the profiles of $S_z$ evaluated for the fields reported in Fig.1,
and we note that the sign of $S_z$ is not constant along the transverse profiles, a region where $S_z <0$ (red portion of the curves) existing around
$\xi=0$. This reversing of the power flow along the transverse profile of the nonlinear guided waves is particularly evident from Fig.2(b) where we
draw the vector field ${\bf S}$ on the plane $(\xi,\zeta)$ for one of the fields of Fig.1. In order to physically grasp and discuss this unusual
effect we recast Eq.(\ref{Poynt}) in the form
\begin{equation}
{\bf S} = \frac{c}{2\beta \sqrt{\epsilon \mu}} D_x E_x^* \hat{\bf e}_z
\end{equation}
\begin{figure}
\includegraphics[width=0.5\textwidth]{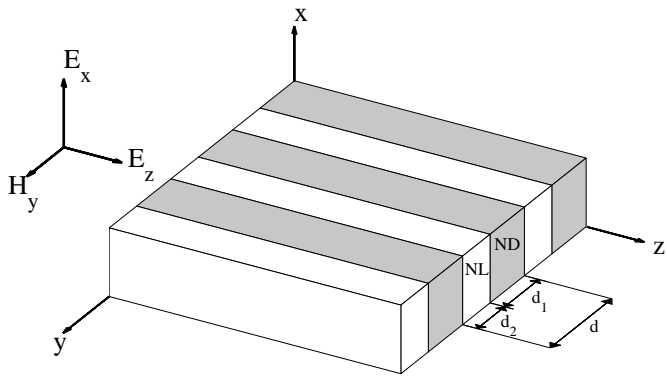}
\caption{Metamaterial layered structure able to support transverse power flow reversing of TM fields, consisting of alternating slabs of a negative
permittivity dielectric (ND) and a standard nonlinear cubic medium (NL).}
\end{figure}
where use of Eqs.(\ref{TM}) and (\ref{consti}) has been made, from which it is evident that the transverse power flow reversing is a consequence of
the sign flipping of $D_x$ along the wave profile while $E_x$ does not change its sign. This implies that, regardless the absolute sign of the
fields, the overall effective dielectric response undergoes a sign reversing due to the fact that, in the first of  Eq.(\ref{consti}), the nonlinear
cubic term can be both smaller and greater than the linear part, depending on the local field strengths. It is worth noting that, although we have
discussed this effect using the considered nonlinear guided waves admitting analytical treatment, the phenomenon is more general and holds for the
wider class of TM fields of the form ${\bf E}(x,z) = e^{i k z} [E_x (x,z) \hat{\bf e}_x + i E_z (x,z) \hat{\bf e}_z]$ since, if $|\partial _x E_x|
\ll k |E_x|$ and $|\partial _x E_z| \ll k |E_z|$ (i.e. the field manly propagates along the $z-$ axis) it is simple to obtain from Maxwell equations
that $S_z = \frac{\omega}{2k} Re[D_x^*(x,z) E_x(x,z)]$ and the transverse power flow reversing can take place through the just discussed mechanism.
We conclude that the predicted power flow reversing is a signature of the extreme nonlinear regime where the cubic nonlinear contribution to the
medium polarizability is not a mere perturbation of the linear part. In this sense the medium behaves as a metamaterial whose character (positive or
negative dielectric constant) locally depends on the field intensity. It is worth stressing that the discussed power flow reversing is very different
from the effect that, in left handed metamaterials, the Poynting vector is antiparallel to the carrier wave vector which is a consequence of the fact
that, in such media, $\epsilon < 0$ and $\mu <0$ (with $n<0$). On the other hand, in our case, $\mu > 0$ and the sign of the power flow is not
uniform being controlled through the field intensity. Note that such an extreme condition can be achieved when the field intensity $|E|^2$ is
comparable or greater than $\epsilon/\chi$, so that, in standard materials where $\epsilon$ is generally of the order of unity and $\chi$ is very
small (of the order of $10^{-20} m^2/V^2$ in semiconductors \cite{Boyddd}), the required intensity is so large to rule out the whole discussed
phenomenology. However, if a metamaterial is employed where $\epsilon$ can be chosen to be much smaller than unity, the intensity threshold can be
reduced to the point of making the extreme nonlinear regime accessible even for intensities much smaller than those employed in standard nonlinear
optics experiments.

Even though the use of a metamaterial ($\epsilon \ll 1$) makes feasible intensities able to trigger the above linear-nonlinear competition, the main
issue remains of finding a medium whose dielectric response is, in the considered intensity range, purely cubic. In fact, the first of
Eqs.(\ref{consti}) is a power series expansion of the constitutive relation $D=D(E)$ in the field strength $E$ and therefore, if the third order is
comparable with the first one, one generally has to consider higher order terms. In order to show that the discussed extreme nonlinear regime can
effectively be achieved, consider the metamaterial structure reported in Fig.3 consisting of alternate linear metamaterial and nonlinear medium
layers of thickness $d_1$ and $d_2$ respectively. The metamaterial is a negative dielectric (ND) whose constitutive relation is ${\bf D} = \epsilon_0
\epsilon_1 {\bf E}$ ($\epsilon_1<0$)  whereas the nonlinear medium (NL) is characterized by the constitutive relation ${\bf D} = \epsilon_0
\epsilon_2 {\bf E} - \epsilon_0 \chi_2 [({\bf E} \cdot {\bf E}^*) {\bf E} + \gamma ({\bf E} \cdot {\bf E}) {\bf E}^* ]$, i.e. it is a standard
isotropic defocusing ($\epsilon_2>0$,$\chi_2>0$) Kerr medium. The media relative permeability are $\mu_1$ and $\mu_2$, respectively. If the spatial
period $d_1 + d_2$ is much smaller than the field vacuum wavelength $2 \pi c / \omega$, the TM field propagating through the structure experiences
the effective response described by Eqs.(\ref{consti}) and characterized by the spatially averaged parameters
\begin{eqnarray} \label{aver}
\epsilon &=& f \epsilon_1 + (1-f) \epsilon_2, \nonumber \\
\chi &=& (1-f) \chi_2, \nonumber \\
\mu &=& \left( \frac{f} {\mu_1} + \frac{1-f}{\mu_2} \right)^{-1}
\end{eqnarray}
where $f=d_1/(d_1+d_2)$ is the fraction of negative dielectric. From these relations, it is evident that suitable values of $\epsilon_1$ and $f$ can
be chosen so that $0 < \epsilon \ll 1$, i.e. the overall medium effective response coincides with the one considered in present Letter. Most
importantly, the medium is able to support the extreme nonlinear regime since if the field is such that $|E|^2 \sim \epsilon/ \chi$, at the same time
one has that $|E|^2 \ll \epsilon_2/ \chi_2$. Therefore the nonlinear medium layers (NL) are in the presence of a field for which their response is
purely cubic and, as a consequence, the overall averaged structure response is purely cubic as well.

Even though the physical origin of the extreme non linear regime is entirely dielectric, magnetic properties of the metamaterial can play an
important role. In fact, from Eqs.(\ref{TM}), Eq.(\ref{Poynt}) and the relations $\xi = \sqrt{\epsilon \mu} (\omega/c) x$, $\zeta = \sqrt{\epsilon
\mu} (\omega/c) z$, we note that $\mu$ plays the role of a scaling parameter for both the field physical sizes and intensity. This implies that, the
field propagating through a medium with $\mu > 1$ can be obtained from the field propagating through a medium with $\mu = 1$ by shrinking its sizes
and reducing its intensity by a factor $1/\sqrt{\mu}$. As a consequence the intensity required for observing the extreme nonlinear regime can be
further reduced exploiting the range of available permeability $\mu$, achievable through specially designed metamaterials.

As a specific example, consider a TM field of wavelength $\lambda = 1.064 \:\: \mu m$ propagating through a layered metamaterial structure for which
$\epsilon_1 = -51.991$ (coinciding, for example, with the real part of the silver permittivity at the considered $\lambda$ \cite{Palikk}), $\mu_1=1$
and $\epsilon_2 = 12.041$, $\mu_2 = 1$, $\chi_2 = 8.742 \times 10^{-19} m^2/V^2$ and $\gamma = 0.5$ (which are the linear and nonlinear parameters
characterizing the GaAs at the considered wavelength \cite{Boyddd,Sheikb}). From Eqs.(\ref{aver}) we obtain, for $f=0.188$, the average effective
permittivity $\epsilon = 0.003$ and the average permeability $\mu=1$. Consider now the nonlinear guided wave whose power flow is reported in Fig.(2)b
which is characterized by $u_{x\infty} = 0.65$ and a transverse dimensionless width $\Delta \xi \simeq 2$ ($\Delta \xi$ also coincides with the width
of the transverse portion of the field where the Poynting vector is antiparallel to the propagation direction). The physical width of the considered
wave is $\Delta x = (\lambda/2\pi) (\Delta \xi / \sqrt{\epsilon}) \simeq 6.2 \: \mu m$ whereas the maximum of its normalized Poynting vector is $S_z
/ S_0 \simeq 0.1$ (see Fig.2) so that the wave is characterized by the intensity $S_z = 0.1 \: S_0 \simeq 3 \: MW/cm^2$. It is worth stressing that
the considered micron-sized confined wave is observable with an intensity ($\sim MW /cm^2$) much smaller than that ($\sim GW/cm^2$) required for
exciting a spatial soliton (of the same width and at the same wavelength) propagating through a GaAs sample \cite{Kivsha}. Referring to the above
example, from a conceptual point of view, if everything is left unchanged but we choose the extreme permeability $\mu_1 = -0.234$, the considered
nonlinear guided wave is characterized by the (sub-wavelength) physical width $\Delta x  \simeq 0.57 \: \mu m$ and the intensity $S_z \simeq 0.27 \:
MW/cm^2$, i.e. they are both reduced roughly by a factor of $10$.

In conclusion, we have shown that nonlinear metamaterial with very small dielectric permittivity can support the propagation of electromagnetic beams
exhibiting transverse power flow reversing, i.e. the Poynting vector changes sign along their transverse profile. Such an unusual phenomenon is one
of the manifestations of the underlying extreme nonlinear regime where the nonlinear contribution to the polarizability can even exceed the linear
contribution, a situation never occurring in general nonlinear optical setups. It is worth stressing that the novel extreme nonlinear regime and its
foreseen manifestations are observable even at relatively small intensities since the metamaterial has a small linear dielectric permittivity. In
addition, we have noted that metamaterials with relevant magnetic properties can allow the observation of the novel regime even for very small
intensities and characterized by a very tight sub-wavelength field confinement. Therefore we conclude that the designing of complex metamaterials
where dielectric nonlinearity combines with a pronounced magnetic response, can play a fundamental role for conceiving sub-wavelength nonlinear
devices.

%%%%%%%%%%%%%%%%%%%%%%%%%%%%%%%%%%%%%%%%%%%%%%%%%%%%%%%%%%%%%%%%%%%%%%%%%%%%%%%%%%

\end{document}